\documentclass[11pt]{article}
\usepackage{hyperref}
\pdfoutput=1
\begin{document}
\title{Electro-Osmotic Instability and Chaos near Ion-Selective Surfaces}
\author{Mathias B. Andersen, Clara L. Druzgalski, Scott M. Davidson,\\
and Ali Mani \\
\\\vspace{6pt} Center for Turbulence Research, \\ Stanford University, Stanford, CA 94305, USA}
\maketitle
\begin{abstract}
We present a "fluid dynamics video" showing results from direct numerical simulations of
electrokinetic transport in a binary electrolyte in two systems: (1) next to an ion-selective membrane and (2)
around a metallic cylinder. The governing equations are the Poisson--Nernst--Planck and
Navier--Stokes and our DNS resolves the time and length scales associated with the thin electric double layers. Our results show transition to chaotic dynamics when the applied voltage
across the characteristic length scale is above a threshold of approximately 0.5 Volt. More details, in
particular regarding system (1), are given in our paper [Druzgalski \emph{et al}. POF \textbf{25}, 110804 (2013)].
\end{abstract}
\section{Description of video}


The video starts by presenting two systems: (1) electrodialysis for water purification and (2) induced-charge electro-osmosis for pumping and mixing in microfluidics. Common to these systems are ion-selective transport driven by either surface conduction or inherent selectivity of the boundary conditions. The resulting electrohydrodynamic flow is chaotic above a threshold voltage of approximately 0.5 Volt as measured over the characteristic length in the system [being the diffusion-layer width in (1) and the cylinder radius in (2)].

The applied voltage is 120 and 50 (in thermal units) and the dimensionless Debye screening length $10^{-4}$ and $10^{-3}$ in system (1) and (2), respectively, and in both systems the material electrohydrodynamic coupling constant is $0.5$. More details are available in Ref. \cite{druzgalski}.

The movie displays colorplots of the salt concentration and free charge density. Visualization of salt concentration is enhanced artificially in system (1), by showing concentration to the power of 1/3 (instead of a linear scale) to make the low concentration region more visible. The movie shows fluctuations of disparate length and time scales with unprecedented resolution and currently not accessible by experimental techniques.

The consequence of the microscopic chaos on the system-level response is significant for both systems. For system (1) it leads to mass transport above the diffusion limit, so-called overlimiting current. In the case of system (2), chaos significantly suppresses the ICEO velocities, in a time-averaged sense. The resulting mean flow can be up to a factor of 4-5 below those predicted by current steady ICEO models. These results can explain puzzling discrepancies between theories and experiments of ICEO.

\end{document}